# Plasmonic nano-aperture label-free imaging (PANORAMA)

Nareg Ohannesian[1], Ibrahim Misbah[1], Steven H. Lin[5], and Wei-Chuan Shih[1,2,3,4,*]

1 Department of Electrical and Computer Engineering, University of Houston, 4800 Calhoun Road, Houston, Texas 77204, United States of America
2 Department of Biomedical Engineering, University of Houston, 4800 Calhoun Road, Houston, Texas 77204, United States of America
3 Department of Chemistry, University of Houston, 4800 Calhoun Road, Houston, Texas 77204, United States of America
4 Program of Materials Science and Engineering, University of Houston, 4800 Calhoun Road, Houston, Texas 77204, United States of America
5 Department of Radiation Oncology, The University of Texas MD Anderson Cancer Center, Houston, Texas, United States of America
*E-mail: wshih@uh.edu

## Abstract

Label-free observation of nanoparticles by far-field optical microscopy is challenging because their ability to scatter or absorb light dramatically diminishes with decreasing size. Surface plasmon resonance (SPR) and localized surface plasmon resonance (LSPR) imaging have shown promises and respective limitations. For instance, it is challenging to achieve diffraction-limited resolution in SPR imaging; its refractive index sensitivity is not as localized as desired. LSPR imaging is typically employed with dark-field microscopy on sparse noble metal nanostructures, leading to low light throughput and incomplete imaging. Here we demonstrate ultra near-field index modulated PlAsmonic NanO-apeRture lAbel-free iMAging (PANORAMA) that addresses existing issues for both SPR and LSPR imaging. PANORAMA produces diffraction-limited lateral resolution with higher surface sensitivity compared to SPR. Its system configuration is identical to a standard bright-field microscope using a trans-illumination tungsten-halogen lamp instead of a laser or other high-intensity light sources. Additionally, PANORAMA addresses the sparse sampling issue in LSPR imaging by achieving dense sampling with a large imaging fill factor. The bright-field approach provides much higher light throughput compared to dark-field microscopy. Overall, our technique can provide a *panoramic* view both laterally and longitudinally – overcoming the lack of imaging depth for both SPR and LSPR imaging and the insufficient lateral sampling for LSPR imaging. We have demonstrated that PANORAMA can size single nanoparticle down to 25 nm, count individual nanoparticles in a cluster, and dynamically monitor single nanoparticle approaching the plasmonic surface down to the millisecond timescale. PANORAMA is potentially useful in single biological nanoparticle analysis of exosomes and viruses.

## Introduction

Dielectric nanoparticles are considered transparent "phase" objects in optical physics, i.e., they contribute a slight variation to the local refractive index. Owing to their minute physical dimension, there are two fundamental issues preventing their facile characterization at single particle level by optical microscopy. First, because their scattering cross-section is extremely small, they can hardly stand out from the background, representing a **detection** challenge. Second, because of their size being much smaller than the optical diffraction limit, resolving individual units among a group of nanoparticles within close proximity is difficult, posing a *resolution* challenge [1]. To overcome the detection challenge, dark-field illumination has been employed while suffering low light throughput thus slow imaging speed [2, 3]. To circumvent the resolution challenge, recent advances in superresolution imaging have gained considerable ground, but are limited to fluorescence microscopy [4]. Non-fluorescent nanoparticles can flow through one by one in flow cytometry so there is no need to resolve [5]. This type of serial measurement is usually in the dark-field mode with limited throughput. To increase throughput, intense light illumination has to be employed which could introduce adverse effects. Using a single pixel detector, they cannot take advantage of the rapid advances in camera technology for high-throughput, high-information content (HIC) imaging. In addition, any flow-through technology can only measure the nanoparticle as it travels by the detection area. This prevents time-lapse monitoring of the same nanoparticles. Besides, once a nanoparticle is measured, it is difficult to retrieve it for any additional measurement or processing. For label-free measurements, e.g., dark-field scattering, there is a lack of molecular information – only particle count and size distribution can be obtained [6]. Label-free optical techniques also include Interferometric Scattering microscopy (iSCAT) and Interferometric Reflectance Imaging Sensor (IRIS). The former has been primarily employed for tracking applications while the latter is based on traditional interference technique so the surface sensitivity is inferior to plasmonic techniques. Nevertheless, both iSCAT and IRIS require an interferometry setup and a high quality laser. Labeled techniques require additional labeling and washing procedures of reporter labels such as fluorescent dyes, quantum dots, and gold nanoparticles, which increases complexity and cost [7]. Fluorescence imaging being the most commonly used labeling technique for biological nanoparticles such as exosomes and viruses, but single nanoparticle detection is still challenging due to factors such as insufficient fluorescence and photobleaching.

Light excited surface plasmons (SP) have become the basis of many optical imaging and sensing techniques [8]. Relying on propagating surface plasmon polariton (SPP), surface plasmon resonance (SPR) is a highly effective mechanism to detect minute refractive index (RI) changes on a gold thin [9]. Because most detection targets have a RI higher than that of the ambient medium (e.g., air or water), SPR can detect surface binding without "labeling" with reporters such as fluorescent dyes, bypassing many potential labeling associated issues. Surface plasmon resonance imaging (SPRI) broadens the repertoire of SPR sensing by incorporating an imaging component [10-12]. However, SPRI typically cannot achieve diffraction-limited lateral resolution with respect to the excitation wavelength due to the wave propagating nature of SPP that "smears" the point spread function. Recently, improved spatial resolution has been demonstrated by interferometric SPRI with image processing [13]. The concept of digital holography has also been implemented to image single nanoparticles by SPR with near-field optics [14]. Longitudinally, the evanescent field of SPP extends into the medium with a 1/e distance ~240 nm, which can be considered as the first order approximation of its sensing range above the gold film [12, 15-17]. A predominant SPRI instrument configuration is based on the Kretschmann design where a prism is employed to provide total internal reflection (TIR) excitation, a required condition for momentum matching. It is worth noting that SPRI can only sense a few hundreds of nanometers above the gold surface, so it requires another imaging modality for the parts of specimen outside of this distance. For example, a standard SRPI system constructed based on an inverted microscope requires two light sources, one to provide the bottom-up TIR excitation, and the other a top-down transmission illumination [17]. If a single camera is employed for detection, it must be time-shared by the two imaging modalities.

Localized surface plasmons (LSP), in contrast, refers to non-propagating SPs localized in metallic nanostructures and nanoparticles with a size comparable or smaller than the excitation wavelength [8].

Similar to SPR, LSPR is also sensitive to the surface RI changes and has been heavily pursued in optical sensing of surface molecular binding [18, 19]. Due to the non-propagating nature of LSP, LSPR imaging (LSPRI) would address a key drawback in SPRI by eliminating the smearing effect, thus making diffraction-limited lateral resolution possible [20]. Longitudinally, LSPRI also features shorter sensing distance from the surface, which can provide better sensitivity toward the surface compared to SPRI [15, 21]. Nevertheless, these tantalizing capabilities have not been fully realized. To date, LSPRI embodiments utilize colloidal nanoparticles or nanostructured substrates as sparse sensing units in the form of dark-field scattering microscopy [21, 22]. In the former case where colloidal nanoparticles are employed, their distribution is not pre-arranged nor controllable [23, 24]. In the latter case, arrays of sparsely distributed nanostructures such as pillars, posts, spikes, etc., have prevented continuous lateral (i.e., in the x-y plane) sampling [25, 26]. Compared to how SPRI is employed, where the gold substrate surface can be 100-percent utilized for continuous sampling, the existing LSPRI can only provide laterally sparse images with small imaging fill factor. Failure to supply continuous sampling has several drawbacks such as lower efficiency, missing spatial context, and "blind" to anything outside the sensing near-field both laterally and longitudinally.

In this paper, we demonstrate ultra near-field index modulated PlAsmonic NanO-apeRture lAbel-free iMAging (PANORAMA) that addresses existing issues for present SPRI and LSPRI techniques. On one hand, PANORAMA can produce diffraction-limited lateral resolution free of the previously mentioned smearing effect in SPRI. PANORAMA also has higher surface sensitivity due to the LSPR decay length being shorter than that of SPR. Its system configuration is identical to a standard bright-field microscope using a trans-illumination tungsten-halogen lamp. Therefore, PANORAMA simultaneously images everything within the microscope objective's depth of focus with a single lamp source and a single camera. However, the intensity ratio increases as the target becomes closer to the imaging substrate. Furthermore, PANORAMA addresses the sparse sampling issue in dark-field LSPRI and provides dense sampling with a large imaging fill factor. In other words, our technique can provide a *panoramic* view both laterally and longitudinally – overcoming the lack of imaging depth for both SPRI and LSPRI, and the insufficient lateral sampling for LSPRI. PANORAMA is demonstrated on a high-density Arrayed Gold Nanodisk on Invisible Substrate (AGNIS) measuring 360 nm in diameter, 50 nm in thickness, and 100 nm in edge-to-edge distance. AGNIS has been fabricated using nanosphere lithography followed by a self-aligned substrate undercut. Partial substrate removal to alter LSPR was demonstrated in the literature [28,29]. However, given the relatively large nanodisk dimensions, our AGNIS features significantly blue-shifted LSPR extinction peak due to both far-field plasmonic coupling [27] and substrate undercut [27-30]. We demonstrate the ability of PANORAMA to image dielectric nanoparticles as small as 25 nm by a standard transmission bright-field microscope with a tungsten-halogen lamp. In addition to ultrahigh sensitivity to deep sub-100 nanoparticles, PANORAMA can also provide their size information. Furthermore, using the arrival time difference in a dynamic imaging mode, individual nanoparticles in a cluster with interparticle distance well below the diffraction limit of the current optical system (330 nm) can be counted. Moreover, the longitudinal distance between a nanoparticle and AGNIS can be monitored using the dynamic imaging mode.

**Materials and Methods**

**Materials**

Polystyrene beads of sizes 25 nm, 50 nm, 100 nm, 200 nm, 300 nm, 460 nm, and 750 nm were purchased from Sigma-Aldrich. Ethanol (200 proof) was purchased from Decon Laboratories, Inc. Gold sputtering target was purchased from ACI Alloys, Inc. Argon gas (99.999%) was used for RF-sputter etching.

**Methods**

**Fabrication of AGNIS.** Fabrication steps involve deposition of 2 nm of Titanium as an adhesion layer and then 80 nm of the gold film using E-beam evaporation. A monolayer of polystyrene beads of average diameter 460 nm was assembled over the gold film. The substrate was exposed to oxygen plasma etching to shrink the size of the polystyrene beads, followed by Argon Ion milling to etch away the uncovered part of the gold. The polystyrene beads were washed away via sonication. This generated a two-dimensional polycrystalline array of gold nanodisks with an average diameter of 360 nm with a pitch size (center-to-center distance) of 460 nm. This gold nanodisk array was immersed in a buffer HF to undercut the glass substrate beneath the disks (Supplementary Fig. S1a-b). The LSPR extinction peak of AGNIS is at 620 nm in air and 690 nm in water (Fig. S1c). The gold nanodisks occupy ~55% of the total substrate footprint. However, this is a conservative estimate because the LSPR electromagnetic field extends further away from the edges of nanodisk. As shown in Fig. 5i, the 1/e distance for lateral electric field decay is ~30 nm, which brings the effective lateral imaging fill factor to ~75%.

**Optical Setup.** White light from a tungsten-halogen lamp passes through a condenser (IX-LWUCD, Olympus) and illuminates the sample on an inverted microscope (IX71, Olympus). The transmitted light passes through an infinity corrected 60X water immersion lens with a 1.2 numerical aperture (UPLSAPO60XW, Olympus). The light exiting the side-port is relayed to an electron multiplied charge coupled device (EMCCD; ProEM 1024, Princeton Instruments) via a 4f system, where a bandpass filter with 650-670 nm passband (FB660-10, Thorlabs) at its Fourier plane (Supplementary Fig. S2).

**Principle of PANORAMA.** The light scattering cross-section ($\sigma_{scatt}$) of a spherical particle is proportional to the 6$^{th}$ power of the particle diameter ($d$) given by $\sigma_{scatt} = (2\pi^5 d^6 n_{med}^4 / 3\lambda_{inc}^4) * |(m^2 - 1)/(m^2 + 2)|$, where $n_{med}$ is the refractive index of the medium surrounding the particle, and $m$ is the ratio of the refractive indices of the particle and medium [1, 5, 13]. Unlike most light scattering techniques, PANORAMA does not solely rely on scattered light to detect nanoparticles. With the trans-illumination geometry, the camera receives transmitted light passing through the AGNIS after a bandpass filter. The bandpass filter is located at near the half-max wavelength on the left shoulder of the AGNIS's LSPR extinction curve (supplementary Fig. S1c). When an imaging target resides outside the AGNIS's longitudinal sensing range, i.e., too far away from the surface, it will show up as a standard light scattering object with its intensity reduced when transmitted through the AGNIS. However, as the imaging target approaches AGNIS, the elevated local RI causes the LSPR extinction curve to red shift. The LSPR red shift causes increased light transmission within the imaging wavelength range and acts as if a virtual nano-aperture is formed right beneath the target. We emphasize that the nano-aperture allows higher transmission for both the nanoparticle scattered light and the unscattered incidence light. A quantitative study can be found in supplementary Fig. S3, where differential light transmission via PANORAMA is compared to the LSPR shift for different particle sizes. However, it is critically important to recognize that the increased transmission of the unscattered incidence light is crucial for detecting weakly scattering nanoparticles (sizes < 200 nm).

**Results and Analysis**

*Nanoparticle detection*

Polystyrene (PS) bead sizes of 750, 460, 300, 200, 100, 50, and 25 nm were used to show PANORAMA's performance in detecting tiny phase objects. A ratiometric image ("intensity ratio") is obtained by dividing the image after nanoparticle settlement by the image without nanoparticles. For a control experiment where a ratiometric image is obtained by dividing two images without nanoparticles, resulting in an image histogram with a mean of 1 and a standard deviation of 0.013, which is used as a baseline intensity ratio (IR) image. A threshold IR value is selected to be 1 + 3*0.013~1.04. Assuming Gaussian statistics, an IR value larger than 1.04 indicates particle detection with a *p-value* < .001. Fig. 1a-g shows PANOMARA images of nanoparticles of all sizes after the thresholding process. To compare PANORAMA's performance with standard bright-field microscopy, all PS bead sizes were imaged on a glass coverslip using the same setup (Fig. 1h-l). PANORAMA provides images of detected beads of all sizes (Fig. 1a-g). In contrast, bright-field images (Fig. 1h-l) show a gradual decrease in intensity toward smaller bead size, and no detection can be made for 100 nm (Fig. 1l) and smaller beads (not shown).

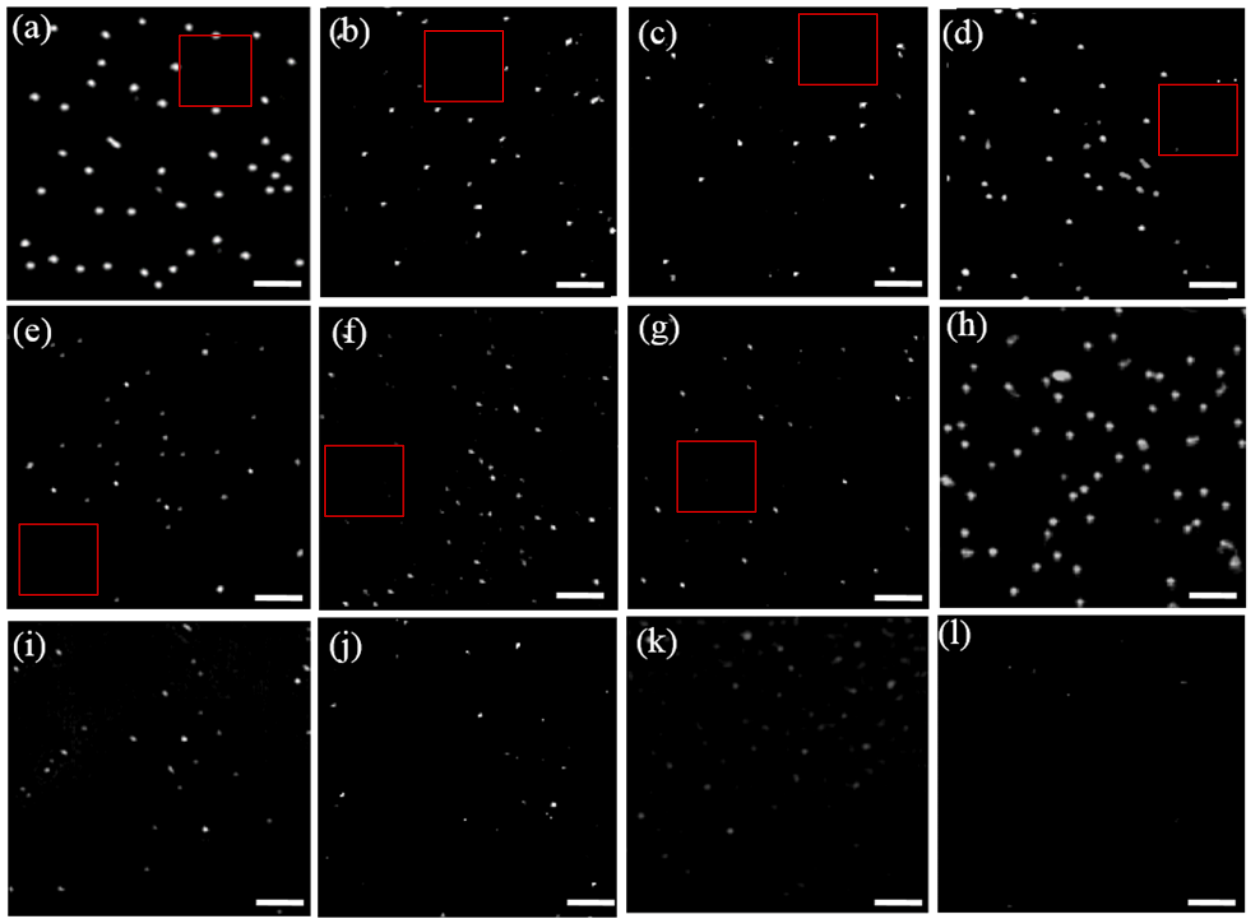

*Fig. 1: PANORAMA images of PS beads: (a) 750 nm, (b) 460 nm, (c) 300 nm, (d) 200 nm, (e) 100 nm, (f) 50 nm and (g) 25 nm. Bright-field images of PS beads: (h) 750 nm, (i) 460 nm, (j) 300 nm, (k) 200 nm, (l) 100 nm. The squared regions are where the background IR was calculated in Fig. 2(l). Scale bar: 5 μm.*

Fig. 2 shows a histogram analysis constructed from IR values recorded from all detected nanoparticles in Fig. 1 and control experiments. By applying a gaussian fit, the mean IR values by PANORAMA were 1.81, 1.33, 1.28, 1.22, 1.175, 1.14, and 1.1 for decreasing nanoparticle size (Fig. 2a-g), while the values from bright-field microscopy were 1.57, 1.17, 1.13, 1.06, 1.02, 1, and 1 (Fig. 2h-k). As mentioned earlier, an IR value smaller than 1.04 is considered not detectable. Considering the smallest PS bead size of 25 nm, PANORAMA provides an IR of 1.1 with sigma 0.014, which is well separated from the background histograms which centers at 1 with sigma 0.0131 (Fig. 2l), suggesting the possibility of detecting even smaller particles. Overall, PANORAMA has been shown to detect PS bead sizes down to 25 nm, whereas bright-field microscopy cannot detect beads equal or smaller than 100 nm.

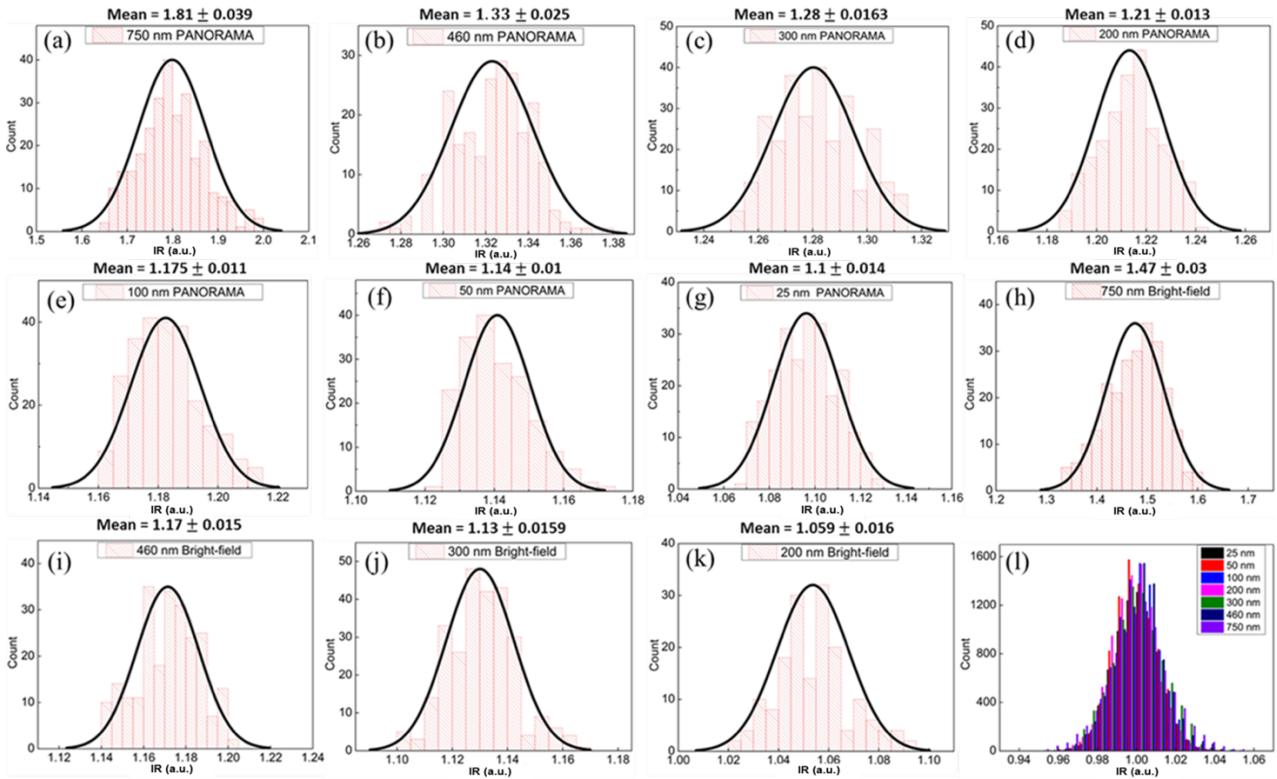

*Fig. 2: Compiled IR histograms with the corresponding mean and standard deviation of PANORAMA images of PS beads: (a) 750 nm, (b) 460 nm, (c) 300 nm, (d) 200 nm, (e) 100 nm, (f) 50 nm and (g) 25 nm. Compiled IR histograms with the corresponding average and standard deviation of bright-field images of PS beads: (h) 750 nm, (i) 460 nm, (j) 300 nm, (k) 200 nm. (l) Corresponding IR histogram of the background from the squared regions in Fig. 1.*

Considering the ability of PANORAMA to detect sizes reaching 25 nm, it cannot spatially resolve nanoparticle sizes below the diffraction limit. However, the mean and standard deviation of the histograms shows a correlation between nanoparticle size and IR where increasing nanoparticle size registers larger IR. Fig. 3 illustrates the image IR vs. nanoparticle size extracted from Fig. 2 for both PANORAMA (black squares) and bright-field microscopy (red circles). It is observed that PANORAMA maintains a larger IR from 11%-22% among various nanoparticles size. It is remarkable to note that PANAROMA's IR continues to decrease for nanoparticle smaller than 300 nm, suggesting that PANORAMA can provide size information beyond the diffraction-limited resolution. In addition, PANORAMA's IR curve does not seem to plateau at the small nanoparticle size tested (25 nm), implying that the limit has not been reached yet.

The results here suggest PANORAMA has better contrast over IRIS for nanoparticle smaller than about 120 nm. For example, PANORAMA provides a contrast of ~14% for 25 nm PS nanoparticles while IRIS has a contrast < 2%. In addition, PANORAMA maintains ~10% contrast for 25 nm PS nanoparticles but IRIS has already reached the noise floor. These results should not be too surprising. It is well known that LSPR has much better surface sensitivity than SPR and traditional interferometry.

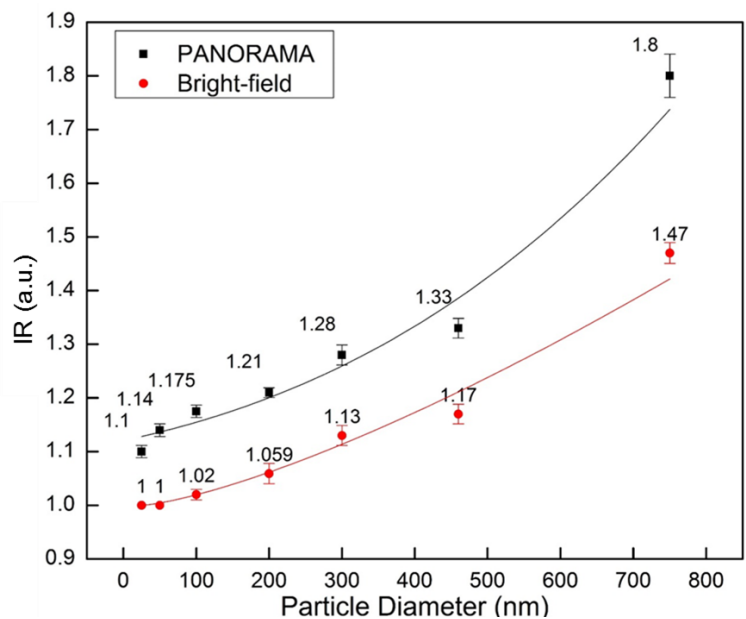

*Fig. 3: IR vs. nanoparticle diameter via PANORAMA (black squares) and by bright-field microscopy (red circles). The curves are provided for visual guidance.*

*PANORAMA nano-aperture size dependence on nanoparticle size*

An interesting property of PANORAMA lies in the nano-aperture size dependence on the nanoparticle size. The smallest achievable aperture size is roughly the gold disk diameter of 360 nm, which is comparable to the diffraction limited resolution (330 nm) and larger than the equivalent EMCCD pixel size (195 nm) at the sample plane. We note that the gold disks can be made smaller than both the diffraction limited resolution and the camera pixel size. Fig. 4a-d shows images of detected 25 nm PS beads, where a single pixel with the highest IR can be found to be accompanied with several surrounding lower IR pixels. The pattern varies according to the specific positions of the gold disk with respect to the EMCCD pixel grid, and to a certain degree the relative position of the PS bead on the gold disk. The minimum pattern size is equivalent to a patch of 2x2 pixels (Fig. 4a), but most detected nanoparticles appear in a patch of 3x3 pixels (Fig. 4c, d). PS bead sizes of 50, 100, 200, and 300 nm show similar patterns as 25 nm PS beads (Fig. 4e-h), suggesting the nanoparticle caused a single gold disk to shift in most cases. When the nanoparticle size becomes comparable to the gold disk, e.g., 300 nm PS beads, the pattern can occasionally appear as a slightly larger patch (Fig. 4i). A plausible explanation is as the nanoparticles become larger, the probability increases for a single nanoparticle to causes LSPR shift for the adjacent disk(s). Nano-aperture across multiple gold disks becomes more apparent for larger nanoparticles as shown in Fig. 4j for 460 nm and Fig.

4k for 750 nm, respectively. The sampling density can be improved by higher magnification and smaller camera pixels, which will be investigated in the future.

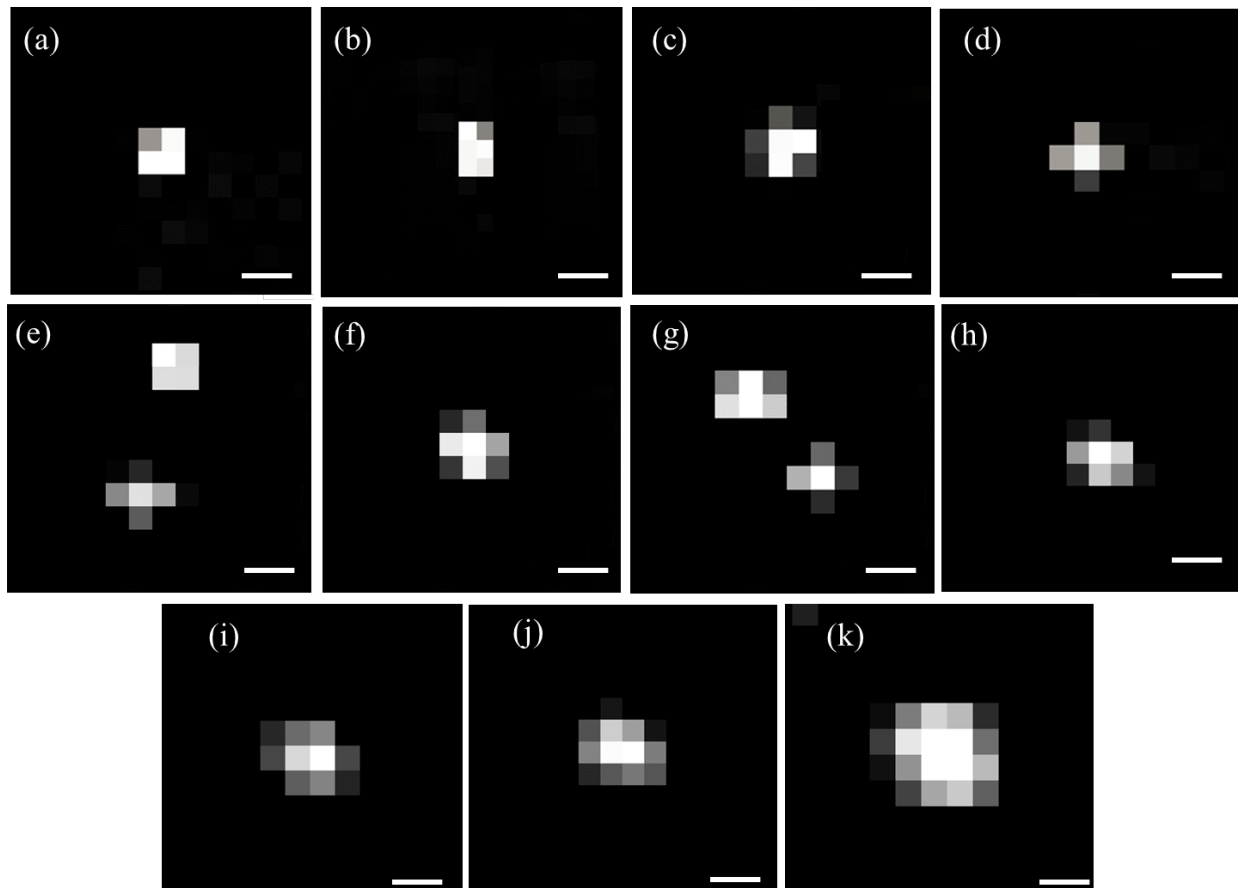

*Fig. 4: Detected particle pixel illumination grid for particle sizes (a-d) 25 nm; (e) 50 nm, (f) 100 nm, (g) 200 nm, (h-i)300 nm; (j) 460 nm; and (k) 750 nm. Scale bar: 390 nm.*

*Interpreting nanoparticle to surface distance by dynamic monitoring*

A key strength of PANORAMA is its outstanding light throughput, which enables dynamic imaging at millisecond timescale. To demonstrate this experimentally, we performed dynamic imaging at 33 Hz frame rate and 10 ms integration time per frame during nanoparticle settling. We note that the frame rate is limited by preventing the camera from overheating. Similar IR values have been obtained from 25 nm PS beads with as little as 1 ms integration time, suggesting the potential of reaching kHz frame rate. Fig. 5a-g shows the IR values of various size PS beads as they settle on the AGNIS. Nanoparticle settling time ($\tau$) can be defined as the travel time between the nanoparticle is first detected and IR plateau (complete settlement) as shown in Fig. 5a-g, suggesting a trend of increased $\tau$ with increased nanoparticle size (Fig. 5h). Assuming the settling speed does not strongly depend on size, the larger $\tau$ for larger nanoparticles suggests that they can be detected farther away from the AGNIS surface. The plateau IR values for all particle sizes agree well with the previously established values after nanoparticles settlement (Fig. 3). To gain more insight on the surface sensitivity of PANORAMA, we performed finite difference time domain (FDTD) simulations where a 50 nm PS nanoparticle was placed at various distances from the AGNIS. As shown in Fig. 5i, the IR of the particle increases as the nanoparticle becomes closer to AGNIS with a 1/e distance of 25 nm. Interpreting the 50 nm nanoparticle result in Fig. 5b in the light of the simulated results, the nanoparticle was outside the detectable distance (~100 nm) from the AGNIS surface prior to t~99 msec, after which it moved toward the AGNIS and arrived at the surface at t~190 msec. As shown previously, 50 nm PS nanoparticle does not provide detectable light scattering in the bright-field imaging mode, suggesting the settling process was imaged entirely by PANAROMA. In contrast, larger nanoparticle (e.g., >100 nm) can provide detectable light scattering as long as the nanoparticle enters the imaging system depth of focus (DOF) ~ +-413 nm. In other words, larger nanoparticles can be first detected when they are significantly farther away from the AGNIS surface entirely due to its own light scattering. In the case of 750 nm nanoparticles (Fig. 5g), it first entered the DOF at t~66 msec, then moved toward the AGNIS surface and arrived at t~333 msec. It is remarkable that PANORAMA contributes additional IR even for larger nanoparticles as shown in Fig. 5g. The time for the 750 nm PS bead to settle on both the glass coverslip and AGNIS's surface is the same, however, the slope of the IR vs. time curve for PANORAMA is significantly higher than that of the standard bright-field image, indicating higher distance sensitivity.

These results suggest that the $\tau$ during nanoparticle settlement can provide an additional means for nanoparticle sizing. They also suggest that it is possible to monitor the distance between the nanoparticle and AGNIS and obtain velocity with a resolution far better than the diffraction limit provided careful calibration and higher imaging speed.

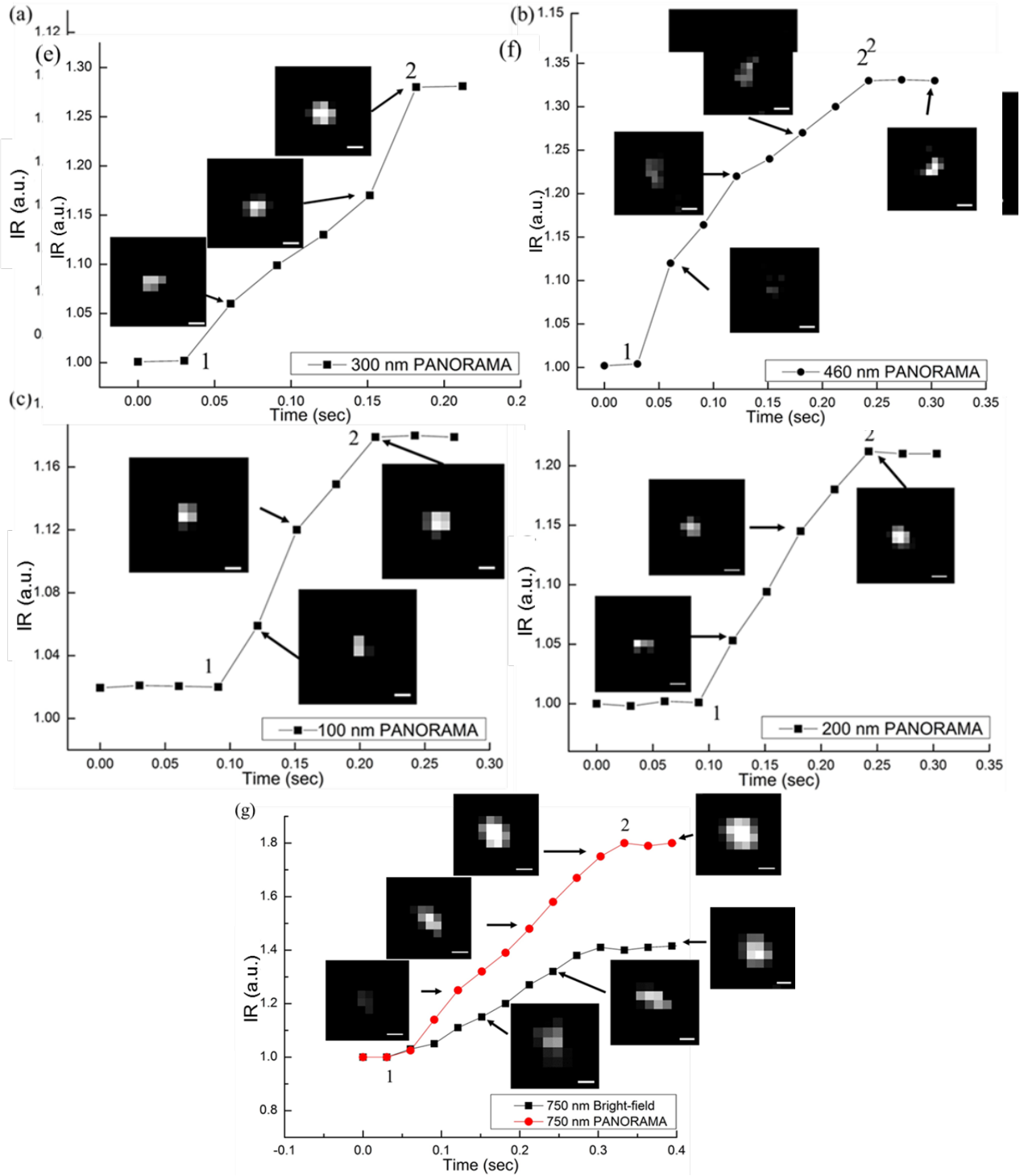

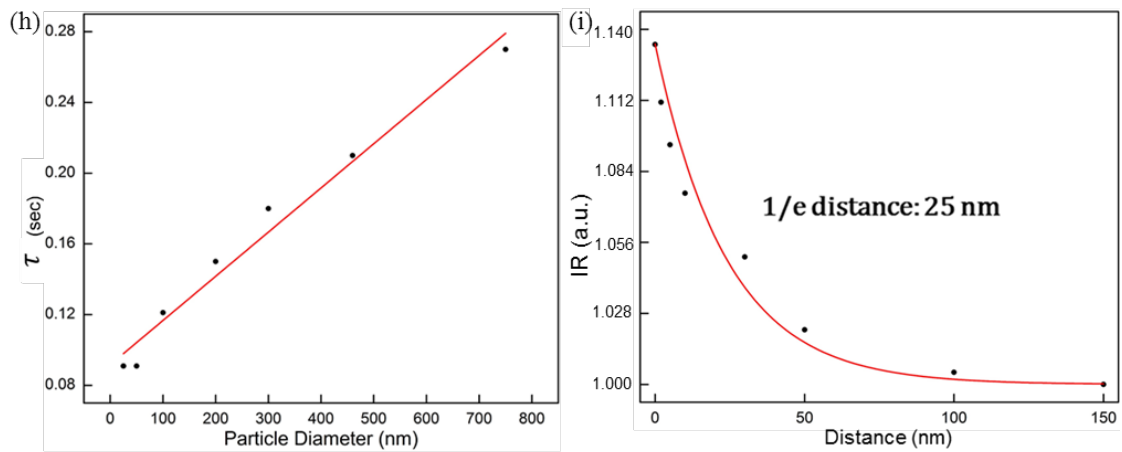

*Fig. 5 PANORAMA IR of (a) 25 nm, (b) 50 nm, (c) 100 nm, (d) 200 nm, (e) 300 nm and (f) 460 nm. (g) The IR of 750 nm PS bead via PANORAMA (red) and via bright-field imaging (black). (h) Particle settlement time (τ) vs. particle size. τ was calculated from the points labeled 1 to 2 for all particle sizes in (a-g). (i) FDTD simulation of IR vs. distance between a 50 nm particle and AGNIS. Scale bar: 390 nm.*

The dynamic imaging capability can be further exploited to count nanoparticle clusters with interparticle distance smaller than the diffraction limit. Counting is achieved by monitoring the IR fluctuations over time. As shown in Fig. 6a where a 25 nm PS bead was initially settled on the surface at time *i*, followed by a second bead settling into the same region at time *ii* and form a cluster (Fig. 6b). The process is evident by monitoring the IR over time in Fig. 6e where the first bead shows a typical IR value for a single 25 nm PS bead. At time *ii*, the IR jump indicates the settlement of a second particle. The same behavior can be seen for 100 nm PS bead in Fig. 6c, d, and f.

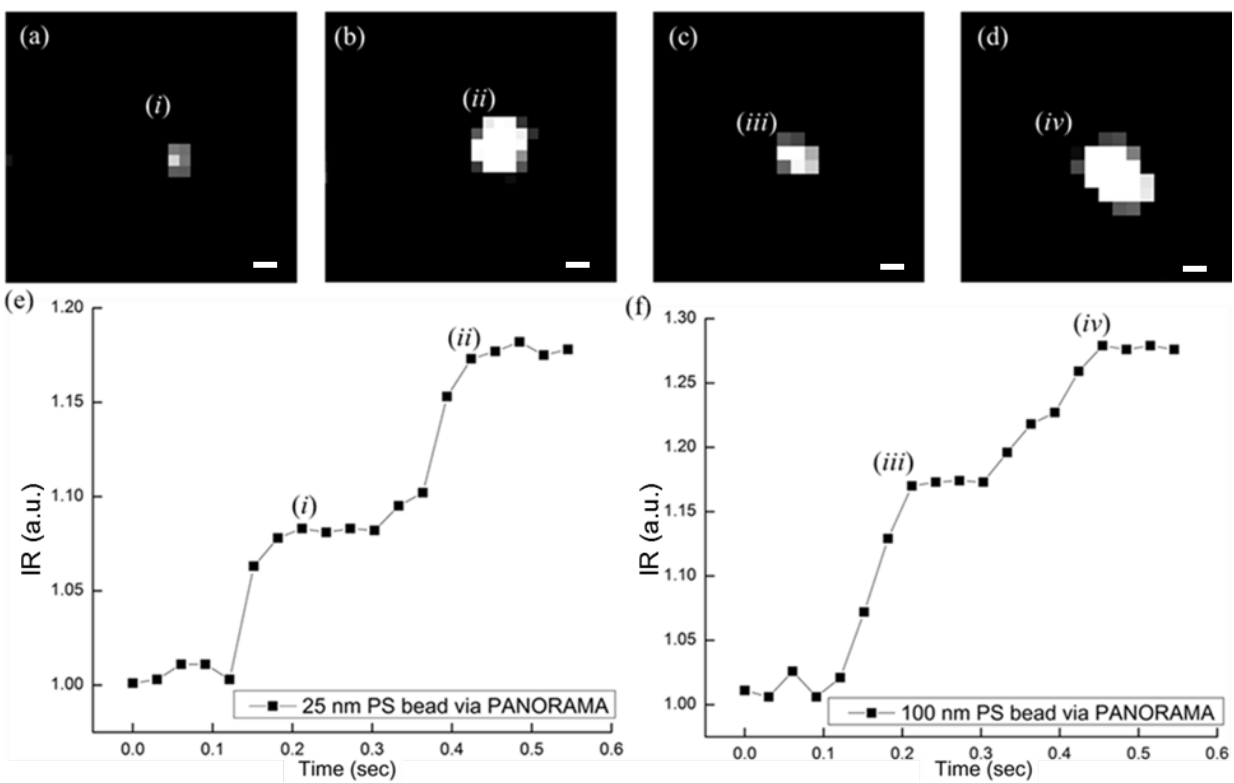

*Fig. 6: (a-b) Detected 25 nm PS bead at time frame i and ii in (e). (c-d) Detected 100 nm PS bead at time frame iii and iv in (f). Dynamic monitoring of IR vs. Time where two (e) 25 nm and (f) 100 nm PS beads settle at the same location. Scale bar: 390 nm.*

**Conclusion**

PlAsmonic NanO-apeRture lAbel-free iMAging (PANORAMA) has been demonstrated as a novel nanoparticle imaging technique. Its basic principle stems from virtual nano-apertures modulated by ultra near-field refractive index. Instead of measuring the light scattered by the nanoparticle, which diminishes dramatically with reducing size, PANORAMA detects both the scattered and unscattered light modulated by the nano-aperture, thus enjoys unprecedented sensitivity. PANORAMA addresses several limitations in existing SPR and LSPR imaging techniques by providing large lateral imaging fill factor and extended longitudinal imaging range while achieving diffraction-limited imaging resolution. Its system configuration is identical to a standard bright-field microscope using a trans-illumination tungsten-halogen lamp instead of lasers or other high-intensity light sources, and a single camera. The bright-field approach provides much higher light throughput for dynamic imaging at the millisecond time scale compared to dark-field microscopy that suffers low light throughput. We have demonstrated that PANORAMA can size single nanoparticle down to 25 nm, dynamically monitor single nanoparticle approaching the plasmonic surface, and count individual nanoparticles in a cluster. PANORAMA would provide new capabilities in label-free imaging and single nanoparticle analysis. More importantly, molecular imaging can be envisioned with a functionalized AGNIS surface for single biological nanoparticle analysis including extracellular vesicles such as exosomes [31] and viruses.

**References**

1. Bohren, C.F. and D.R. Huffman, *Absorption and scattering of light by small particles*. 2008: John Wiley & Sons.
2. Fu, Y.H., et al., *Directional visible light scattering by silicon nanoparticles.* Nature communications, 2013. **4**(1): p. 1-6.
3. Wagner, T., H.-G. Lipinski, and M. Wiemann, *Dark field nanoparticle tracking analysis for size characterization of plasmonic and non-plasmonic particles.* Journal of nanoparticle research, 2014. **16**(5): p. 2419.
4. Betzig, E., et al., *Imaging intracellular fluorescent proteins at nanometer resolution.* Science, 2006. **313**(5793): p. 1642-1645.
5. Zhu, S., et al., *Light-scattering detection below the level of single fluorescent molecules for high-resolution characterization of functional nanoparticles.* ACS nano, 2014. **8**(10): p. 10998-11006.
6. Wang, F., et al., *Scattered light imaging enables real-time monitoring of label-free nanoparticles and fluorescent biomolecules in live cells.* Journal of the American Chemical Society, 2019. **141**(36): p. 14043-14047.
7. Jin, D., et al., *Nanoparticles for super-resolution microscopy and single-molecule tracking.* Nature methods, 2018. **15**(6): p. 415-423.
8. Maier, S.A., *Plasmonics: fundamentals and applications*. 2007: Springer Science & Business Media.
9. Raether, H., *Surface plasmons.* Springer Tracts in Modern Physics, 1988. **111**: p. 1.
10. Maley, A.M., et al., *Characterizing single polymeric and protein nanoparticles with surface plasmon resonance imaging measurements.* ACS nano, 2017. **11**(7): p. 7447-7456.

11. Shan, X., et al., *Imaging the electrocatalytic activity of single nanoparticles.* Nature nanotechnology, 2012. **7**(10): p. 668-672.
12. Brolo, A.G., *Plasmonics for future biosensors.* Nature Photonics, 2012. **6**(11): p. 709.
13. Yang, Y., et al., *Interferometric plasmonic imaging and detection of single exosomes.* Proceedings of the National Academy of Sciences, 2018. **115**(41): p. 10275-10280.
14. Nelson, J.W., et al., *Digital plasmonic holography.* Light: Science & Applications, 2018. **7**(1): p. 1-11.
15. Otte, M.A., et al., *Identification of the optimal spectral region for plasmonic and nanoplasmonic sensing.* ACS nano, 2010. **4**(1): p. 349-357.
16. Shan, X., et al., *Measuring surface charge density and particle height using surface plasmon resonance technique.* Analytical chemistry, 2010. **82**(1): p. 234-240.
17. Shan, X., et al., *Imaging local electrochemical current via surface plasmon resonance.* Science, 2010. **327**(5971): p. 1363-1366.
18. Willets, K.A. and R.P. Van Duyne, *Localized surface plasmon resonance spectroscopy and sensing.* Annu. Rev. Phys. Chem., 2007. **58**: p. 267-297.
19. Sepúlveda, B., et al., *LSPR-based nanobiosensors.* nano today, 2009. **4**(3): p. 244-251.
20. Fritzsche, J., et al., *Single particle nanoplasmonic sensing in individual nanofluidic channels.* Nano letters, 2016. **16**(12): p. 7857-7864.
21. Chen, P., et al., *Multiplex serum cytokine immunoassay using nanoplasmonic biosensor microarrays.* ACS nano, 2015. **9**(4): p. 4173-4181.
22. Liang, K., et al., *Nanoplasmonic quantification of tumour-derived extracellular vesicles in plasma microsamples for diagnosis and treatment monitoring.* Nature biomedical engineering, 2017. **1**(4): p. 1-11.
23. Kawawaki, T., et al., *Potential-scanning localized plasmon sensing with single and coupled gold nanorods.* The journal of physical chemistry letters, 2017. **8**(15): p. 3637-3641.
24. Howes, P.D., R. Chandrawati, and M.M. Stevens, *Colloidal nanoparticles as advanced biological sensors.* Science, 2014. **346**(6205): p. 1247390.
25. Raghu, D., et al., *Nanoplasmonic pillars engineered for single exosome detection.* PloS one, 2018. **13**(8): p. e0202773.
26. Shen, B., et al., *Plasmonic nanostructures through DNA-assisted lithography.* Science advances, 2018. **4**(2): p. eaap8978.
27. Zhao, F., et al., *Far-field plasmonic coupling in 2-dimensional polycrystalline plasmonic arrays enables wide tunability with low-cost nanofabrication.* Nanoscale Horizons, 2017. **2**(5): p. 267-276.
28. Dmitriev, A., et al., *Enhanced nanoplasmonic optical sensors with reduced substrate effect.* Nano letters, 2008. **8**(11): p. 3893-3898.
29. Aćimović, S.S., et al., *Superior LSPR substrates based on electromagnetic decoupling for on-a-chip high-throughput label-free biosensing.* Light: Science & Applications, 2017. **6**(8): p. e17042-e17042.
30. Misbah, I. and W.-C. Shih. *Plasmonic Sensors on Invisible Substrates*. in *Bio-Optics: Design and Application*. 2019. Optical Society of America.
31. Ohannesian, N., et al., *Commercial and emerging technologies for cancer diagnosis and prognosis based on exosomal biomarkers.* Journal of Physics: Photonics, 2020.

## Author contributions

W.S. conceived the idea and directed the study. N.O. and W.S. designed the experiment and analyzed the data. N.O. performed the experiment and produced the figures. I.M. fabricated the AGNIS and performed

FDTD simulations. N.O., S.H.L., and W.S. provided substantial input to the project and to the writing of the manuscript.